\documentclass[12pt]{article}

\usepackage{graphics,epsfig}
\begin{document}
\title{Modeling Rumors:\\ The No Plane Pentagon French Hoax Case}
\author{Serge Galam
\\Laboratoire des Milieux D\'esordonn\'es et H\'et\'erog\`enes\thanks 
{Laboratoire
associ\'e au CNRS (UMR n$^{\circ}$ 7603),  galam@ccr.jussieu.fr},\\
Case 86, 4 place Jussieu, 75252 Paris Cedex 05, France}

\date{To appear in Physica A}
\maketitle

\begin{abstract}

The recent astonishing wide adhesion of french people to the rumor claiming
`No plane did crash on the Pentagon on September the 11", is given a 
generic explanation
in terms of a model of minority opinion spreading. Using a majority rule
reaction-diffusion dynamics, a rumor is shown to invade for sure a social group
provided it fulfills simultaneously two criteria. First it must initiate with a
support beyond some critical threshold which however, turns out to be 
always very low.
Then it has to be consistent with some larger collective social paradigm of the
group. Othewise it just dies out. Both conditions were satisfied in 
the french case
with the associated book sold at more than 200 000 copies in just a 
few days. The
rumor was stopped by the firm stand of most newspaper editors stating 
it is nonsense.
Such an incredible social dynamics is shown to result naturally from 
an open and free
public debate among friends and colleagues. Each one searching for 
the truth sincerely
on a free will basis and without individual biases. The polarization 
process appears
also to be very quick in agreement with reality. It is a very strong 
anti-democratic
reversal of opinion although made quite democratically. The model may 
apply to a large
range of rumors.

\end{abstract}

\newpage

\section{Introduction}

Very recently, the assertion from an individual stating indeed there 
were no crash
plane on the Pentagon on September the 11, received in France an unprecedented
massive adhesion notwithstanding the obvious nonsense of the assertion.
Within a few days more than two hundred thousands copies of his book 
[1] were sold.
Every one was debating the issue with millions of people adhering to 
the lie. To stop
this overloading of misinformation, all newspaper leader-editors made 
firm stand on
denouncing unanimously an ashamed and unacceptable make-up of reality [2]. A
counter book with a detailed proof of the Pentagon attack was even 
published [3].

But since then, all has been forgotten, or almost. No one is any 
longer interested
in the issue. But yet this astonishing event may prove useful to 
grasp the complex
dynamics behind the more general and broad phenomenon denoted under 
the generic name
of rumor [4]. It offers an opportunity to analyze the process of 
individual choice
making from public and open discussions. In particular it allows to connect
the effect of backmind collective social paradigms in yielding the direction
of a public opinion polarization.

The subject of rumor formation is becoming of a strategic importance 
at all levels
of society. The control and possible handling to manipulate information are now
major issues in social organizations including economy, politics, 
defense, fashion,
and even personal affairs. Especially with the existence of Internet 
which provides a
support to anybody to say anything and then consequently to be 
possibly heard by
millions of people. To be read can imply to be automatically 
perceived like truth, and
retransmitted as such to others. There exist no parapets.

However, information shared by a very great number of people does not 
obviously prove
of anything its authenticity. But it can induce quite concrete and 
sometimes dangerous
follow up acts. It may also happen that once a point of view on some specific
issue is widely adopted, the presentation of objective facts proving 
its falseness,
does not produce the abandonment from this same false point of view. 
At contrast, a
rumor can prove to be true while first set false by official media. 
The frontier
between a rumor and information turns out to be very fragile [4].

To try to put on some new light on this rather complicated 
phenomenon, we evoke a
recent study on minority spreading in random geometries [5]. Using a majority
rule reaction - diffusion model, its shows how an opinion at the 
extremely minority
beginning propagates in a random geometry of social meetings. It is 
found to always
gain an overwhelmed majority in a group provided it starts beyond a 
certain very
low threshold value [5, 6], if it is also coherent with some social paradigm.
Otherwise it dies out. The associated dynamics appears to be 
extremely quick (few days)
in both cases of total spreading of dying out in accordance with 
empirical fact about
rumor phenomena [4].

It is worth to stress that above model is not the reality itself. But 
it aims by
making crude approximations at discovering certain essential and 
radical aspects of
this very reality which are otherwise totally hidden by the 
complexity of the full
phenomenon.  Such a sociophysics treatment [7] of a social problem is 
symptomatic of a
new emerging trend from physicists of disorder [8-13].

The paper is organized as follows. The chronology and the
content of the Pentagon french hoax are first reviewed in Section 2. 
In Section 3 we
present the minority model [5]. It is then applied to the french case 
in Section 4. The
massive and quick adhesion of french people to the no plane Pentagon 
hoax is shown to
result from an existing collective anti-American bias which is 
independent of the
issue itself [14]. The same mechanism explains why the hoax did not 
spread in other
country like for instance England. To conclude, the existence of 
systematic collective
bias active in the forming of public opinion is emphasized in last Section.

\section{The Pentagon french hoax}

On September the 11 all french media as like all other world media 
announces the news "a
plane has crashed on the Pentagon" in the series of the terrorist 
attacks on the US. The
fact is naturally perceived as an objective truth. No one questioned 
it? There were no
doubt what so ever about the fact itself. Nevertheless, its reality 
could have proven
disturbing for some people, as far as their global ideological 
worldview was concerned.
For those who hold America as a satanic and very powerful country, 
this barbarian and
unacceptable aggression against the same America deeply disturbed 
their global vision
of the world. They had to live with it.

But then, when later on an isolated individual starts diffusing on 
the Net his counter
truth, "not only has no plane crashed on the Pentagon, but moreover 
the blow was
assembled by the United States", all above unease people absorbed 
this counter truth at
once and literally like a saving truth. For them, America was indeed the
beforehand well-presented monster. This coming back to coherent 
ideological world view
certainly acts on tens of thousands of French people. The selling of 
more than 200 000
copies of the hoax claiming book in less tahan few days demonstrated 
such an immediate
release for a huge amount of french people.

However, even if up to twenty percent of the french population was 
immediately adhering
to the lie, its immense majority, that is to say eighty percent, was 
felt not concerned
with this "revelation". For them, it was at best perceived more like 
a sectarian wild
imagining. The phenomenon remained contained and confidential, though 
with a hard core
of believers.

But afterwards, the TV came in to mediate the issue. It has played a 
key role in the
following warming up  process of tuning on a generalized public 
debate. There, one of
the major national french TV channels presented at a large audience 
show the untruth as
a new possible scenario to explain the Pentagon destruction. The 
thesis of no plane
crash was defended together with the claim it was set up by american 
secret services.
This presentation was not put on as the truth but as an alternative 
to the current view
on the event. De facto, it created a doubt in the public mind. From there, the
questioning of the event was legitimated at least as a doubt about 
the nature of what
did really happen to the Pentagon on September the 11.

In addition, the sounding right of the question "why they are no 
pictures of the carcass
plane on the Pentagon", drove an unbearable doubt. The hugeness of 
this revelation made
it a necessity for the people to clear up the issue at stake. 
Consequently, a public
debate started  as the essential medium to resolve the mystery. 
Moreover, as the
response to the absence of plane carcass was counter-intuitive, once 
someone made up
its mind from discussing with friends or colleagues, it could always 
been shaken again
in its view. The fragility of the individual making choice has 
resulted in a series of
local and repeated discussions. The truth was perceived as emerging 
from the making of a
collective truth setting up the facts within a clear explanation. It 
was up to the
public opinion to decide what had actually occurred at the Pentagon 
on September the
11.

Nevertheless, it could have been expected that starting with a 
majority of eighty
percent of the population holding on the truth "a plane has crashed on the
Pentagon", the public debate would automatically lead to enforce the 
truth on the
initial twenty percent of people believing on the untruth "no crush 
plane on the
Pentagon, it was set up by american secret services". But in fact, 
and in an astonishing
manner, the opposite did happen. The lying minority did turn on its 
side the majority
of the people first holding on the truth.

\section{The minority spreading model}

To understand above paradox, let us follow the process of an 
individual searching for
the  truth from open and repeated discussions with friends and 
colleagues. Discussing
this kind of issue occurs at social gatherings at which people chat 
freely about any
matter like the weather, a sport event or some news. These gatherings 
take place at
most at social times like coffee breaks, lunches, or dinners. At each 
one of them, a
small number of people get together, usually from two to six or a bit 
more, to enjoy a
drink or some food. There, while discussing, arguing and drinking, 
often the whole small
group lines up within a more or less consensual opinion [4, 5]. 
However this opinion is
fragile since resulting from an informal discussion and not from an irrefutable
demonstration. As such it is suitable for a shift at another meeting. 
People have no
individual bias towards the issue.

To visualize the phenomenon in a simple manner, we consider a
perfect society where each individual has only one and even power of 
conviction,
whether it is for or against the truth. To be more perfect we also make the
assumption that each individual taken in a local discussion 
eventually aligns along the
position of the initial majority within the actual group. Thus, from 
each group sitting,
informal discussion leads to a local consensus with each participant 
sharing the same
opinion, that is the one of the initial majority. After the dinner, 
lunch or drink,
everyone is convinced of either the truth or the untruth. That is 
because people
are sincere and open mind in their search to answering the question 
of what did happen
to the Pentagon on September the 11. They hold no a priori.

To moderate this local majority rule dynamics,  we introduce the 
possibility for a
group to doubt about the issue. In our model such doubting
states result spontaneously from even groups with an initial local 
parity between the
two opposite opinions. In this case, while doubting from their 
respective individual
arguments, the people need some extra information to establish a choice.  That
additional ingredient is naturally sought in the shared collective social
paradigms which are specific to the overall population the group members
belonged to. These common cultural settings are upstream of any particular
consideration. They are a common sensitivity about wide view of the 
world. Here, for the
french population taken as a whole, it spurs from a rather skeptical 
feeling about
America. Thus in case of a doubt, the group chooses to believe in the 
untruth since it
is coherent to its common background of suspicion towards the United 
States [14]. An
illustration of the dynamical process is showed in Figure 1.

\begin{figure}
\epsfxsize=\columnwidth
\begin{center}
\centerline{\epsfysize=15cm\epsfxsize=11cm\epsfbox{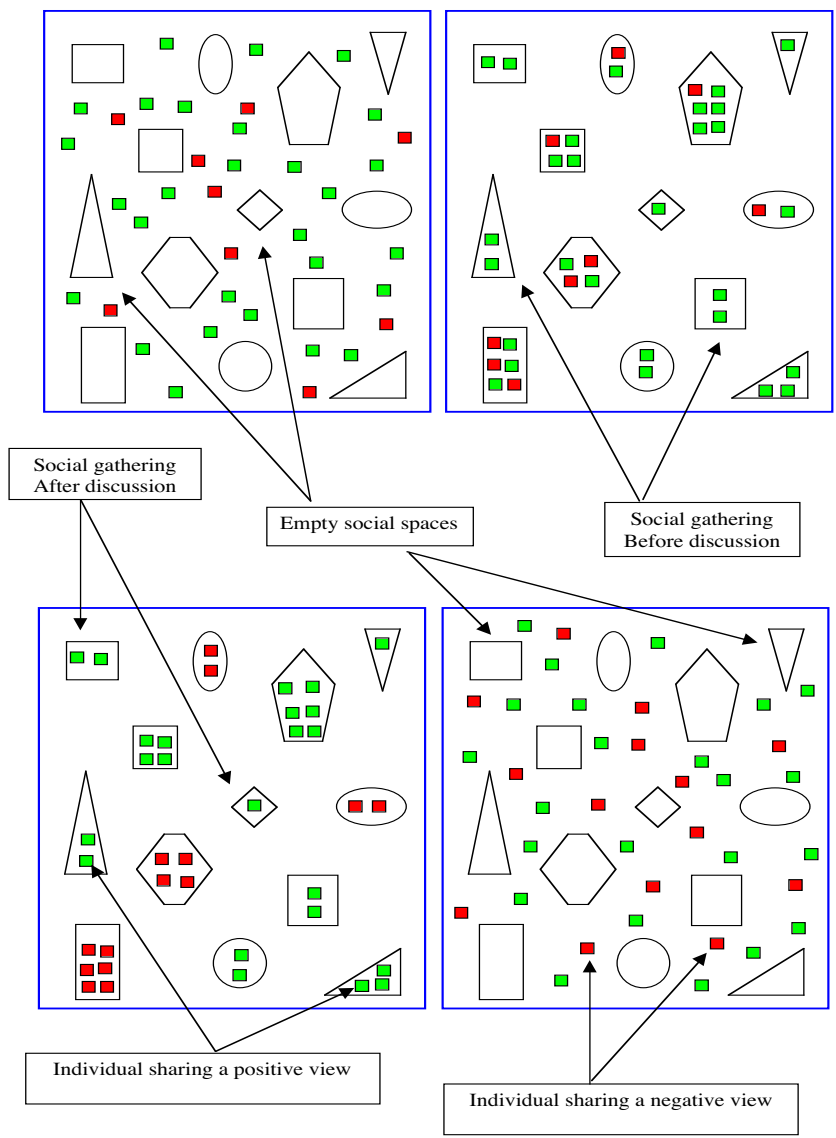}}
\end{center}
\vspace*{-1cm}\caption{A one step social gathering dynamics. Up left, 
people sharing the
two opinions are moving around. Grey are for and black are against. 
No discussion
is occurring with 28 grey and 9 black. Upper right, people are
having lunch by groups of various sizes from one to six. They start discussing.
Nooen yet change its mind. Below left, people are ending their lunch. 
Consensus has been
reached within each group. As a result, they are now 23 grey and 14 
black. Below right,
people are again moving around with no discussion. The balance stays 
at 23/14. }

\end{figure}

It is worth to stress that in another country this collective backmind can be
different. For instance in England it is a rather American 
sympathizing feeling. Thus,
there the same group in the same doubt would decide to believe in the 
reality of the
crash. As is seen below, that is why in England the rumor just died out.

Within the framework of our model, to have a quantitative grasp on the
discussion driven evolution of the respective proportions of people 
holding on the
truth and the untruth, it is necessary to fix ratios for the various 
social-meeting
sizes. Denoting $\{a_i\}$ the probability to be sitting at a group of 
size $i$, we have
the constraint,

\begin{equation}
\sum_{i=1}^L a_i=1 ,
\end{equation}
where $i=1, 2, ..., L$.  The including of one-person groups makes the 
assumption
everyone is gathering simultaneously realistic.

Starting at time $t$ from a $N$ person population, prior to the public
debate everyone is holding an opinion. There are  $N_+(t)$ 
individuals believing to
the truth ``A plane did crash on the Pentagon on September the 11", 
leaving $N_-(t)$
persons sharing the untruth ``No plane crashed on the Pentagon", with 
$N_+(t)+N_-(t)=N$.
Therefore the probabilities to hold respectively on the truth or the 
untruth are,

\begin{equation}
P_{+}(t)= \frac{N_{+}(t)}{N}  ,
\end{equation}
and,
\begin{equation}
P_-(t)=1-P_+(t) .
\end{equation}

 From this initial configuration, people start discussing the issue at 
the first social
meeting.  Each new cycle of multi-size discussions is marked by a 
time increment $+1$.
 From above simple assumption of a majority rule dynamics, with a bias 
in favor of the
untruth in case of a local doubt, at time $(t+1)$ we get for the 
density truth support,

\begin{equation}
P_{+}(t+1)=\sum_{k=1}^L a_k \sum_{j=N[\frac{k}{2}+1]}^k C_j^k P_+(t)^j
\{1-P_+(t)\}^{(k-j)} ,
\end{equation}
where $C_j^k\equiv  \frac{k!}{(k-j)! j!}$ and
$N[\frac{k}{2}+1]\equiv IntegerPart \ of \ (\frac{k}{2}+1)$.\\

In the course of time, the same people will meet again and again
randomly in the same cluster configuration of size groups (see Fig. 
1). At each new
encounter they discuss locally the issue at stake and may change 
their mind according
to above majority rule applied to each local group. To follow the 
time evolution of the
truth support, Eq. (4) is iterated again and again. A monotonic flow 
is obtained
towards either one of two stable fixed points $P_{0}=0$ and 
$P_{1}=1$. The flow and
its direction are  produced by an
unstable fixed point $P_{K}$ located between $P_{0}$ and $P_{1}$. Its value
depends on both the $\{a_i\}$ and $L$.  We denote it the Killing Point.

For $P_+(t)<P_{K}$  it exists a number $n$  such that  $P_+(t+n)=P_{0}=0$
while for $P_+(t)>P_{K}$  it is another number $m$  which yields 
$P_+(t+m)=P_{1}=0$.
It is either a ``Big Yes"
to the truth at $P_{1}=1$ or a ``Big No" to it at $P_{0}=0$.
Both $n$ and $m$  measure the
required time at reaching a stable and final public opinion. Their 
values depend on the
$\{a_i\}$, $L$  and the initial value $P_+(t)$ .
Accordingly, public opinion is found to be non volatile. It 
stabilizes rather quickly
($n$ and $m$ are usually small numbers) to a clear stand towards the 
issue at stake.

\section{Quantitative illustration}

Figure 2 shows the variation of $P_{+}(t+1)$ as function of $P_{+}(t)$ for one
particular sets of the $\{a_i\}$ with $a_1=0$, $a_2=a_3=a_4=1/3$ and 
$a_5=...=a_L=0$.
There
$P_{K}=0.847$ which puts the required initial support to the truth to 
survive the
public debate, at a such very high value of more than $85\%$.
Simultaneously an initial minority above $15\%$ to support the 
untruth is enough to
produce a final total blindness towards the truth. It is a very 
strong reversing
anti-democratic dynamics of opinion although made quite democratically.

\begin{figure}
\epsfxsize=\columnwidth
\begin{center}
\centerline{\epsfbox{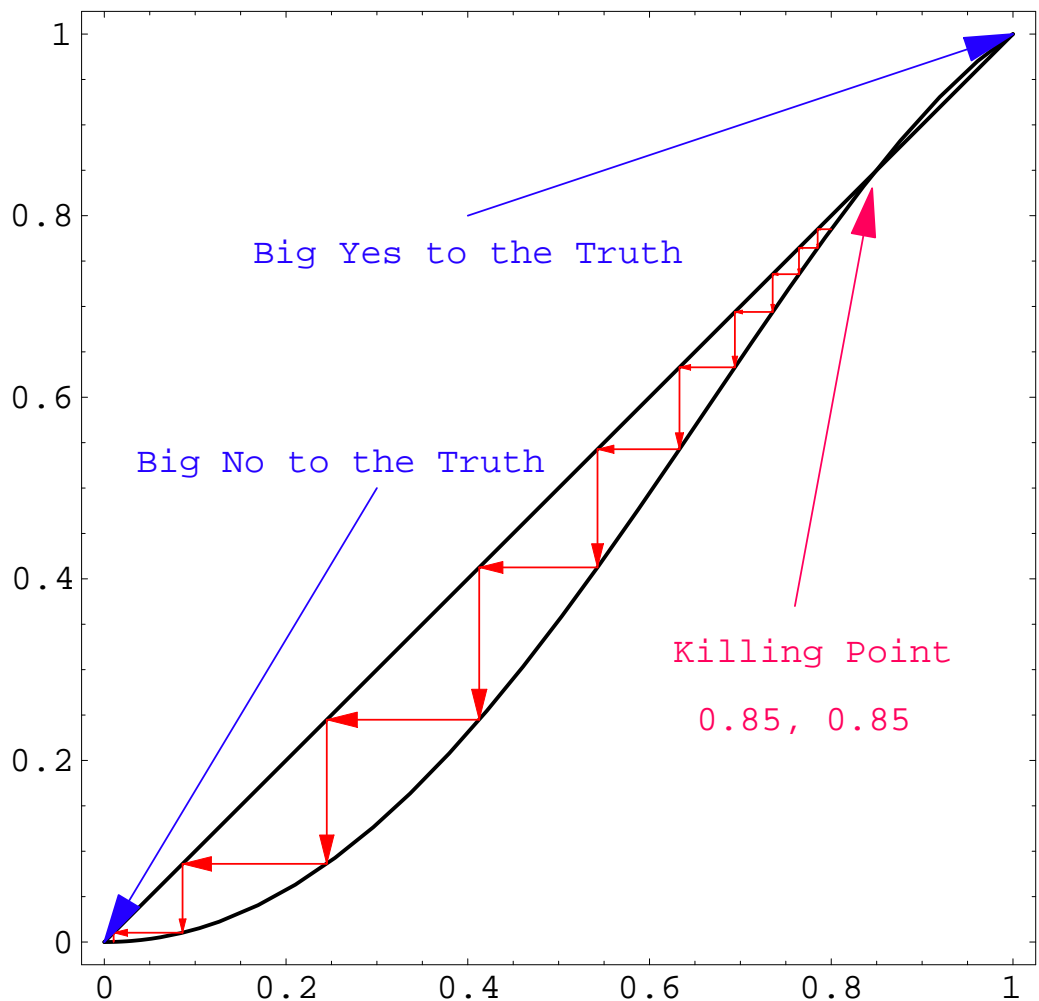}}
\caption{Variation of $P_{+}(t+1)$ as function of $P_{+}(t)$ for
the set $a_1=0$, $a_2=a_3=a_4=\frac{1}{3}$ and $a_5=...=a_L=0$. There 
$P_{K}=0.847$.
Arrows show the direction of the flow.}
\end{center}
\end{figure}

To be more quantitative in above self-blinding dynamics let us consider
above ratio setting with an initial $P_{+}(t)=0.80$  at time  $t$. 
The associated
series in time is  $P_{+}(t+1)=0.78$,
$P_{+}(t+2)=0.77$, $P_{+}(t+3)=0.73$, $P_{+}(t+4)=0.69$,
$P_{+}(t+5)=0.63$, $P_{+}(t+6)=0.54$, $P_{+}(t+7)=0.41$, $P_{+}(t+8)=0.25$,
$P_{+}(t+9)=0.09$, $P_{+}(t+10)=0.01$ and eventually 
$P_{+}(t+11)=0.00$. Eleven cycles
of social local discussions have been enough to turn an initial
$80\%$ of the population supporting the truth, toward an adhesion to 
the untruth. They
just merge quitely and freely with the initial $20\%$ of people who 
first believed to
the untruth. Taking a basis of one discussion a day on average, less 
than  two weeks are
enough to a total crystallization of the lie against the obvious 
truth.  Moreover a
majority favoring the lie is obtained already within six days.

Changing a bit the parameters will change both the Killing Point value and the
number of discussion cycles but yet preserving the basic asymmetry 
and velocity of the
process. Figure 3 shows the number of required discussion cycles to 
get an initial
$30\%$ of layers to turn along their lie the $70\%$ of the population 
who first was
convinced of the truth.

\begin{figure}
\begin{center}
\centerline{\epsfbox{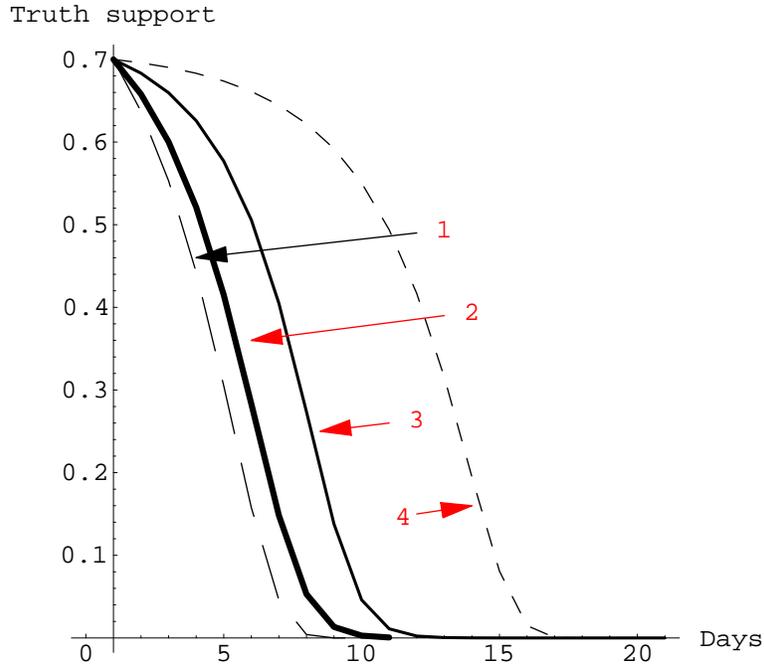}}
\caption{Variation $P_{+}(t)$ as function of successive days with $L=6$.
The initial value  at $t=1$ is $P_{+}(1)=0.70$.
Long dashed line (1): $a_1=0$, $a_2=\frac{1}{2}$, $a_3=\frac{1}{2}$, 
$a_4=a_5=a_6=0$
with $P_{K}=1$.
Heavy thick line (2): $a_1=0.2$, $a_2=0.3$, $a_3=0.2$, $a_4=0.2$, $a_5=0.1$
and $a_6=0$ with $P_{K}=0.85$.
Other line (3): $a_1=a_2=a_3=a_4=0.2$, $a_5=a_6=0.1$. There $P_{K}=0.74$.
Dashed line (4): $a_1=0$, $a_2=0.3$, $a_3=0.7$, $a_4=a_5=a_6=0$ with 
$P_{K}=0.71$.}
\end{center}
\end{figure}

\section{Conclusion}

At this stage it is worth to stress that in real life situations not every
person is open to a mind change. Some fractions of the population will keep
on their opinion whatever happens. Including this effect in the model will not
change qualitatively the results. It will make the polarization 
process not total
with the two stable fixed points shifted towards respectively larger 
and smaller
values than zero and one.

It is also of interest to note that the doubting local state can yield on the
opposite view. For instance in the case of England, with a reversed 
cultural skew
towards America, it is the twenty percent of the lie supporters which 
would have join in
the initial majority of truth supporters, if the debate would had 
been initiated.

Obviously, in reality, not every French person shares the skeptical 
American feeling we
mentioned, and everyone does not change opinion with each social 
meeting. But at the
same time, a rumor does not need to reach hundred percent of the 
population to become
dangerous. In addition, other choices of ratios, for the proportions 
of the various
sizes of the social meetings, would give other figures, but the tendency to
self-propagation of the lie would remain the same as long as the 
initial minority
exceeds a certain value threshold which is nevertheless always low in 
particualar due to
the existence of pair meetings.

We have revealed here tendencies in the dynamics of forming opinion, and
not an exact quantitative determination of any data. It is the 
phenomenon itself,
which must challenge us, more than the figures themselves. We have shown how
individual choices, hold from repeated open discussions with friends 
and colleagues make
the collective public opinion to align rather quickly along some 
social paradigm
hidden a priori commonly shared by the group.

It can be instrumental to note that once launched, such a rumor propagation can
be stopped by non compromise institutional interventions. In the 
example we took, it
has been the solid and firm intervention of most newspaper 
leader-editorialists, which
put an end to the process of reversing an obvious truth. In the case of the
Holocaust deniers, it is the law which made it.

In conclusion, when a rumor starts to develop, it shows the existence 
at a majority of
people of a cultural skewed a priori in the direction that underlies the rumor.
Therefore to avoid wrong and dangerous decisions, it is of a central 
importance to
question the apparent good sense of the social democratic debate. In 
particular to keep
in mind the illusionary character of an individual choice driven from 
open discussions,
can reveal essential in  preserving a country from collective misbehavior.
The model may generalize to a large spectrum of past rumors which did 
happened in
various countries in the world.

\end{document}